\documentclass[twocolumn,showpacs,amsmath,amsfonts,amssymb,superscriptaddress,nofootinbib]{revtex4-1}

\usepackage[utf8]{inputenc}

\pdfoutput=1
\usepackage[usenames,dvipsnames]{color}
\usepackage[colorlinks=true, linkcolor=BrickRed, citecolor=Blue, urlcolor=Blue, filecolor=Blue]{hyperref}
\usepackage{longtable,epsfig,graphicx,verbatim,xspace,multirow,ulem,hhline,placeins,array}

\newcommand{\svee}{{\langle \sigma v \rangle}_{e^+ e^-}\xspace}
\newcommand{\sv}{{\langle \sigma v \rangle}\xspace}

\newcommand{\lcdm}{$\Lambda$CDM\xspace}
\newcommand{\Planck}{{\it Planck}\xspace}
\newcommand{\Neff}{N_{\mathrm{eff}}\xspace}
\newcommand{\mx}{m_{\rm DM}\xspace}
\newcolumntype{K}[1]{>{\centering\arraybackslash}p{#1}}

\begin{document}

\title{Ruling out the light WIMP explanation of the galactic 511 keV line}

\author{Ryan J. Wilkinson}
\affiliation{Institute for Particle Physics Phenomenology (IPPP), Durham University, Durham, DH1 3LE, UK}
\author{Aaron C. Vincent}
\affiliation{Institute for Particle Physics Phenomenology (IPPP), Durham University, Durham, DH1 3LE, UK}
\affiliation{Department of Physics, Blackett Laboratory, Imperial College London, London, SW7 2AZ, UK}
\author{C\'eline B\oe hm}
\affiliation{Institute for Particle Physics Phenomenology (IPPP), Durham University, Durham, DH1 3LE, UK}
\affiliation{LAPTH, U.~de Savoie, CNRS, BP 110, 74941 Annecy-Le-Vieux, France}
\author{Christopher McCabe}
\affiliation{GRAPPA Centre of Excellence, University of Amsterdam, 1098 XH Amsterdam, Netherlands}


\begin{abstract}
Over the past few decades, an anomalous 511 keV gamma-ray line has been observed from the centre of the Milky Way. Dark matter (DM) in the form of light ($\lesssim 10$ MeV) WIMPs annihilating into electron--positron pairs has been one of the leading hypotheses of the observed emission. Given the small required cross section, $ \langle \sigma v \rangle \sim 10^{-30}$ cm$^3~$s$^{-1}$, a further coupling to lighter particles is required to produce the correct relic density. Here, we derive constraints from the \Planck satellite on light WIMPs that were in equilibrium with either the neutrino or electron sector in the early universe. For the neutrino sector, we obtain a lower bound on the WIMP mass of 4 MeV for a real scalar and 10 MeV for a Dirac fermion DM particle, at 95\% CL. For the electron sector, we find even stronger bounds of  7 MeV and 11 MeV, respectively. Using these results, we show that, in the absence of additional ingredients such as dark radiation, the light thermally-produced WIMP explanation of the 511 keV excess is strongly disfavoured by the latest cosmological data. This suggests an unknown astrophysical or more exotic DM source of the signal.
\end{abstract}

\preprint{}

\date{\today}

\maketitle

\section{Introduction}

The emission of a 511 keV gamma-ray line from a spherically symmetric region around the galactic centre has been observed by many experiments over more than four decades~\cite{1978ApJ...225L..11L, 1981A&A....94..214A, 1986ApJ...302..459L, 1988ApJ...326..717S, 1993ApJ...413L..85P, 1997ApJ...491..725P}. By 2003, INTEGRAL/SPI observations had demonstrated that this line originates from the decay of positronium atoms into two photons~\cite{Churazov:2004as, Knodlseder:2005yq, Jean:2005af, Weidenspointner:2006nua}. While this is indicative of an injection of low-energy positrons in the inner kiloparsec of the Milky Way, the signal is uncorrelated with known astrophysical sources. In addition to the bulge, an extended disk-like structure is also seen. However, it is likely associated with radioactive $\beta$-decay of heavy elements produced in stars of the Milky Way disk.

Recently, an analysis of the 11-year data from INTEGRAL/SPI was carried out \cite{Siegert:2015knp}. After a decade of exposure, the significance of the bulge signal has risen to $56 \sigma$, while the disk significance is now $12 \sigma$ in a maximum likelihood fit. New data allow the collaboration to distinguish a broad bulge ($\mathrm{FWHM}_{\rm BB} = 20.55^\circ$) and an off-centre narrow bulge ($\mathrm{FWHM}_{\rm NB} = 5.75^\circ$). There is also significant evidence (5$\sigma$) of a point source at the location of the Sgr~A* black hole near the galactic centre, with a line intensity that is about 10\% of the total bulge (BB + NB) flux. Interestingly, greater exposure of the disk has revealed lower surface-brightness regions, leading to a more modest bulge-to-disk (B/D) ratio of 0.59, compared with previous results (B/D $\sim 1 - 3$).

Low mass X-ray binaries~\cite{Weidenspointner:2008zz}, pulsars and radioactive isotopes produced from stars, novae and supernovae~\cite{Prantzos:2010wi} can yield positrons in the correct energy range for the bulge signal. However, these processes should yield a 511 keV morphology that is correlated with their progenitors' location. For instance, the $\beta^+$ decay of $^{26}$\!Al produced in massive stars also yields a line at 1809 keV, which has been measured by INTEGRAL/SPI \cite{Diehl:2005py}. As expected, this line is not at all correlated with the galactic centre 511 keV emission, although it allows up to 70\% of the positronium formation in the galactic disk to be explained~\cite{Vincent:2012an}. Additionally, estimates of production and escape rates in stars and supernovae suggest that $^{44}$Ti and $^{56}$Ni $\beta$-decay can account for most of the remaining emissivity in the disk \cite{Knodlseder:2005yq, Prantzos:2010wi}. Finally, higher energy sources such as pulsars, magnetars and cosmic ray processes produce $e^\pm$ pairs in the bulge at relativistic energies. However, this would leave a distinct spectral shape above 511 keV, in conflict with the observed spectrum~\cite{Prantzos:2010wi}. The fact therefore remains that the high luminosity of the total bulge emission is not explained by known mechanisms. 

The similarity between the spherically symmetric, cuspy shape of the central bulge emission and the expected galactic dark matter (DM) distribution is highly suggestive of a DM origin. Consequently, an interpretation in terms of self-annihilation of DM has been favoured for some time\footnote{The spatial morphology disfavours a decaying DM origin \cite{Ascasibar:2005rw,Vincent:2012an}.} \cite{Boehm:2002yz, Boehm:2003bt, Ascasibar:2005rw, Vincent:2012an,Cline:2012yx,Chan:2015gbo}. The thermal production of DM through annihilation (as in the WIMP paradigm\footnote{Here, we take the classic definition of a WIMP as a particle that has weak-scale interactions with at least some of the Standard Model particles.}) implies ongoing self-annihilation today. Light DM particles (with a mass $\mx \lesssim 7~ \rm{MeV}$) can produce electron-positron pairs at low enough energies to explain the positronium annihilation signal, while avoiding the overproduction of gamma-rays~\cite{Ascasibar:2005rw, Beacom:2005qv,Sizun:2007ds}. Initial studies could also reproduce the spatial shape of the excess with the standard NFW profile. Later, it was shown that the less cuspy Einasto DM profile yields a significantly better fit to the 511 keV line morphology. In fact, the Einasto shape gives a better fit to the 8-year data than the NB+BB model, with fewer free parameters \cite{Vincent:2012an}. 

The velocity-averaged annihilation cross section required to explain the observed 511 keV flux is $\svee \sim 10^{-30}$ cm$^3$ s$^{-1}$. However, a thermally-produced DM particle requires a cross section at freeze-out $\sv \simeq 3 \times 10^{-26}$ cm$^3$ s$^{-1}$. The two scenarios that satisfy both requirements are:
\begin{enumerate}
\item { {\bf Neutrino ($\nu$) sector}: a dominant annihilation cross section into neutrinos $\sv_{\nu \nu} \simeq 3 \times 10^{-26}$ cm$^3$ s$^{-1}$ at freeze-out.}

\item { {\bf Electron ($e^\pm$) sector}: a velocity-dependent (p-wave) annihilation cross section into electrons $\svee = a + bv^2$, where $bv^2 \simeq 3 \times 10^{-26}$ cm$^3$ s$^{-1}$ dominates at freeze-out.}
\end{enumerate}
In this work, we show that these scenarios are strongly disfavoured by available cosmological data. We begin by presenting their respective impacts on cosmological observables, from the epoch of big bang nucleosynthesis (BBN), recombination and the dark ages. We then show that the latest cosmic microwave background (CMB) data and determinations of the primordial abundances rule out the light WIMP explanation of the 511 keV line.

\section{Neutrino Sector Thermal Production}
\label{sec:neutrinos}

Thermal freeze-out requires annihilation into species lighter than the DM particles. In the case of light DM (below the muon mass), this leaves three channels: electrons, photons or neutrinos. Annihilations into electrons and photons are highly constrained by gamma-ray \cite{Boehm:2002yz} and CMB \cite{Zhang:2007zzh, Galli:2009zc, Slatyer:2009yq, Kanzaki:2009hf, Hisano:2011dc, Hutsi:2011vx,Galli, Finkbeiner:2011dx, 2013PhRvD..87l3513S, Galli:2013dna, Lopez-Honorez:2013cua, Madhavacheril:2013cna, Diamanti:2013bia, Slatyer:2015jla, Kawasaki:2015peu} observations. We therefore first consider the scenario in which the relic density originates via the neutrino channel and the subdominant annihilation rate into $e^\pm$ explains the 511 keV line.

\subsection{BBN and Recombination}

DM annihilations into neutrinos can increase the entropy in the neutrino sector if the DM particles are lighter than $ \sim 15$ MeV and annihilate after the standard neutrino decoupling at $T_{{\rm dec},\nu} \simeq 2.3$ MeV \cite{Kolb:1986nf, Serpico:2004nm, Boehm:2012gr,Ho:2012ug,Steigman:2013yua,Boehm:2013jpa,Nollett:2013pwa,Nollett:2014lwa, Steigman:2014pfa}. This increased energy density is parameterised in terms of the effective number of neutrino species $N_{\rm eff}$. A larger neutrino energy density increases the expansion rate of the universe. If this occurs during BBN, the neutron-to-proton ratio freezes out earlier, leading to an increase in the primordial helium abundance $Y_{\rm P}$ and deuterium-to-hydrogen (D/H) ratio. 

\begin{figure}[ht!]
\centering
\includegraphics[width=0.345\textwidth,angle=-90,clip,trim=0ex 2ex 2ex 0ex]{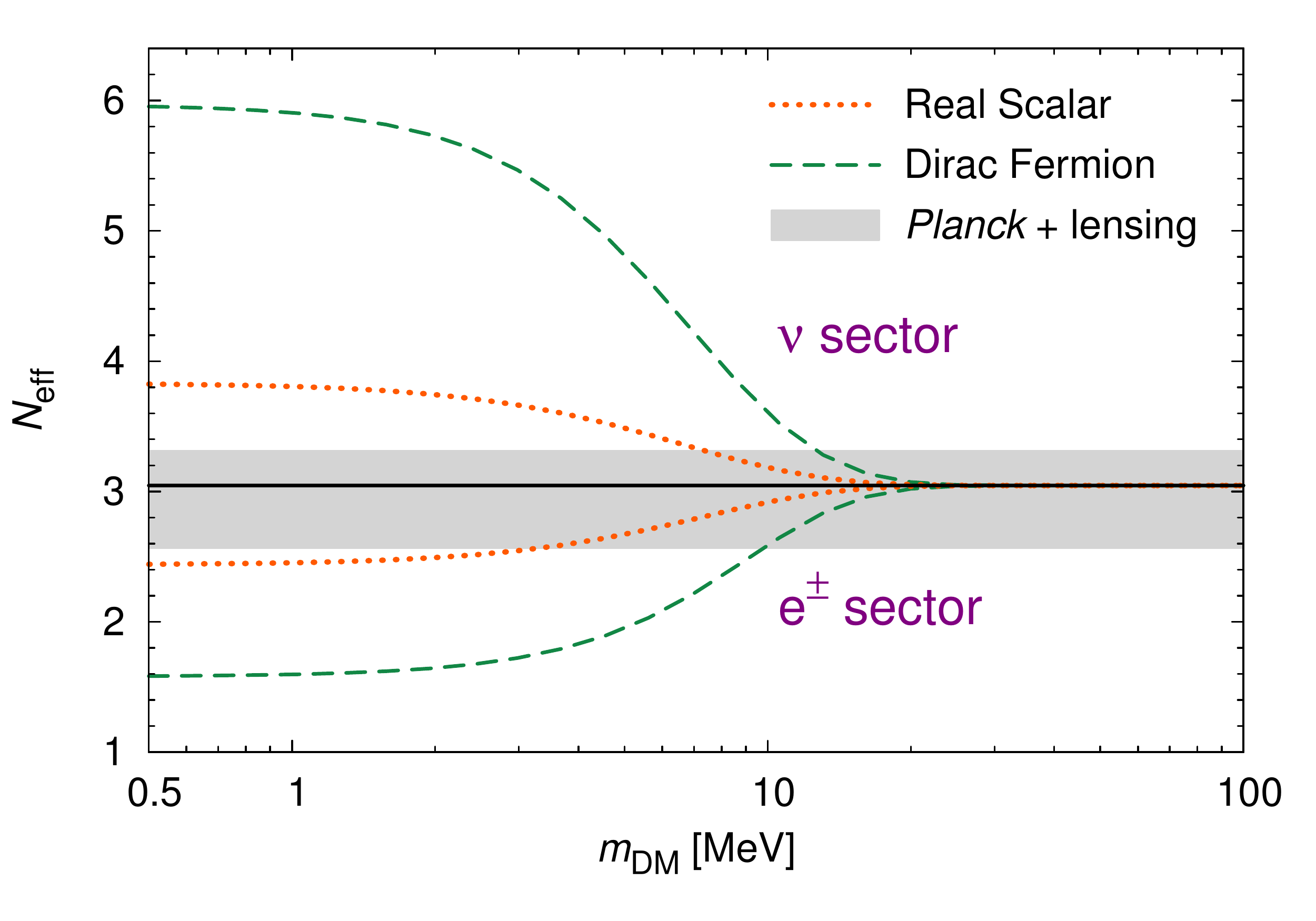}
\caption{The number of relativistic degrees of freedom $N_{\rm eff}$ at the CMB epoch as a function of the DM mass $m_{\rm DM}$ for a real scalar (orange, dotted) and Dirac fermion (green, dashed). For neutrino sector thermal production, the enhancement of $N_{\rm eff}$ is a result of DM annihilations reheating the neutrino sector, as described by Eq.~\eqref{eq:neffcmb}. For electron sector production, the suppression of $N_{\rm eff}$ is due to DM annihilations into $e^+ e^-$ reheating the photon sector, as described by Eq.~\eqref{eq:neffcmb2}. The solid black line corresponds to the standard value of 3.046. Also shown is the 95\% CL favoured region of $N_{\rm eff}$ from the \Planck + lensing dataset (grey band) assuming \lcdm, i.e. $N_{\rm eff} = 2.94 \pm 0.38$~\cite{Planck:2015xua}. Note that a complete MCMC analysis is required to derive constraints from such modifications to $N_{\rm eff}$ as there are well-known degeneracies with the other cosmological parameters.}
\label{fig:neff}
\end{figure}

The same mechanism leads to additional energy in the radiation sector during recombination, again parameterised via $\Neff$. At such low temperatures ($\mx \gg T$),
\begin{equation}
N_{\rm eff}^{{\rm Equil,} \nu} \simeq 3.046 {\left[1 + \frac{g_{\rm DM}}{2}\frac{F(y_\nu|_{T_{{\rm dec},\nu}})}{3.046} \right]}^{4/3}~,
\label{eq:neffcmb}
\end{equation}
where 
\begin{equation}
F(y) = \frac{30}{7\pi^4} \int_{y}^\infty d\xi~\frac{(4\xi^2 -y^2) \sqrt{\xi^2-y^2}}{e^{\xi} \pm 1}~,
\label{eq:Fofy}
\end{equation} 
$g_{\rm DM}$ is the number of internal degrees of freedom for DM and $y_\nu|_{T_{{\rm dec},\nu}} \equiv \mx/T_{{\rm dec},\nu}$~\cite{Boehm:2012gr}. The dependence of $N_{\rm eff}$ on the DM mass for two types of DM particle is illustrated in Fig.~\ref{fig:neff}. This enhances the effect of Silk damping and compounds the impact of a higher $Y_{\rm P}$ in reducing power in the tail of the CMB angular power spectrum. 

Furthermore, the scattering of DM particles with neutrinos during recombination can erase perturbations on small scales due to the process of ``collisional damping''~\cite{Boehm:2000gq,Boehm:2001hm,Boehm:2004th,Bertschinger:2006nq}. It also prevents the neutrinos from free-streaming as efficiently, thus enhancing the CMB acoustic peaks~\cite{Mangano:2006mp,Serra:2009uu,
Wilkinson:2014ksa,Aarssen:2012fx,Farzan:2014gza,Boehm:2014vja,Cherry:2014xra,Bertoni:2014mva,
Schewtschenko:2014fca,Davis:2015rza,Escudero:2015yka,Ali-Haimoud:2015pwa,Schewtschenko:2015rno}. To account for DM--neutrino scattering, the coupled Euler equations that govern the evolution of the DM and neutrino fluid perturbations $\delta_{{\rm DM}/\nu}$ and their gradients $\theta_{{\rm DM}/\nu}$ must be modified to include interaction terms $\propto \sigma_{{\rm DM}-\nu}~(\theta_{\rm DM} - \theta_\nu)$, where $\sigma_{{\rm DM}-\nu}$ is the elastic scattering cross section. The shear $\sigma_\nu$ and higher multipole perturbations $F_{\nu, \ell}$ of the neutrino fluid also acquire terms proportional to $\sigma_{{\rm DM}-\nu}$. These equations and the formalism to modify the Boltzmann code {\sc class}~\cite{Lesgourgues:2011re} are described in Refs.~\cite{Wilkinson:2014ksa,Escudero:2015yka}.

\subsection{The Dark Ages}

Independently of the neutrino sector, the subdominant s-wave annihilations into $e^+e^-$ that produce the galactic 511 keV signal also have strong, observable consequences during the dark ages between the epochs of recombination and reionisation. These effects are measurable in the CMB angular power spectrum. 

Extra electromagnetic energy ionises the intergalactic medium (IGM). This ionisation rescatters CMB photons, leading to a broader surface of last scattering, which suppresses temperature and polarisation correlations on small scales (large multipoles). Enhanced polarisation correlation on large scales is also expected from Thomson scattering at late times. The latest measurements from the \Planck satellite~\cite{Planck:2015xua} set the strongest constraints on energy-injection from DM to date. 

At a given redshift $z$, electromagnetic energy $E$ is injected into the IGM at a rate per unit volume $V$:
\begin{equation}
\frac{{\rm d}E}{{\rm d}t \ {\rm d}V} = f_{\rm eff}(\mx) \ \rho_c^2 \ (1+z)^6 \ \Omega_{\rm DM}^2 \ \zeta \ \frac{\svee}{\mx}~,
\label{eq:dfdt}
\end{equation}
where $\rho_c$ is the critical density, and $f_{\rm eff}(\mx)$ is the effective efficiency of energy deposition into heating and ionisation, weighted over redshift. The latest determination of $f_{\rm eff}$ can be found in Refs. \cite{Slatyer:2015kla, Slatyer:2015jla}. Constraints on Eq.~\eqref{eq:dfdt} are usually quoted in terms of the redshift-independent quantity $p_{\rm ann} \equiv f_{\rm eff}(\mx)\svee/\mx$. Finally, $\zeta =1$ when the DM and its antiparticle are identical, and 1/2 otherwise.

\begin{figure}[ht!]
\centering
\includegraphics[width=0.345\textwidth,angle=-90,clip,trim=0ex 2ex 2ex 0ex]{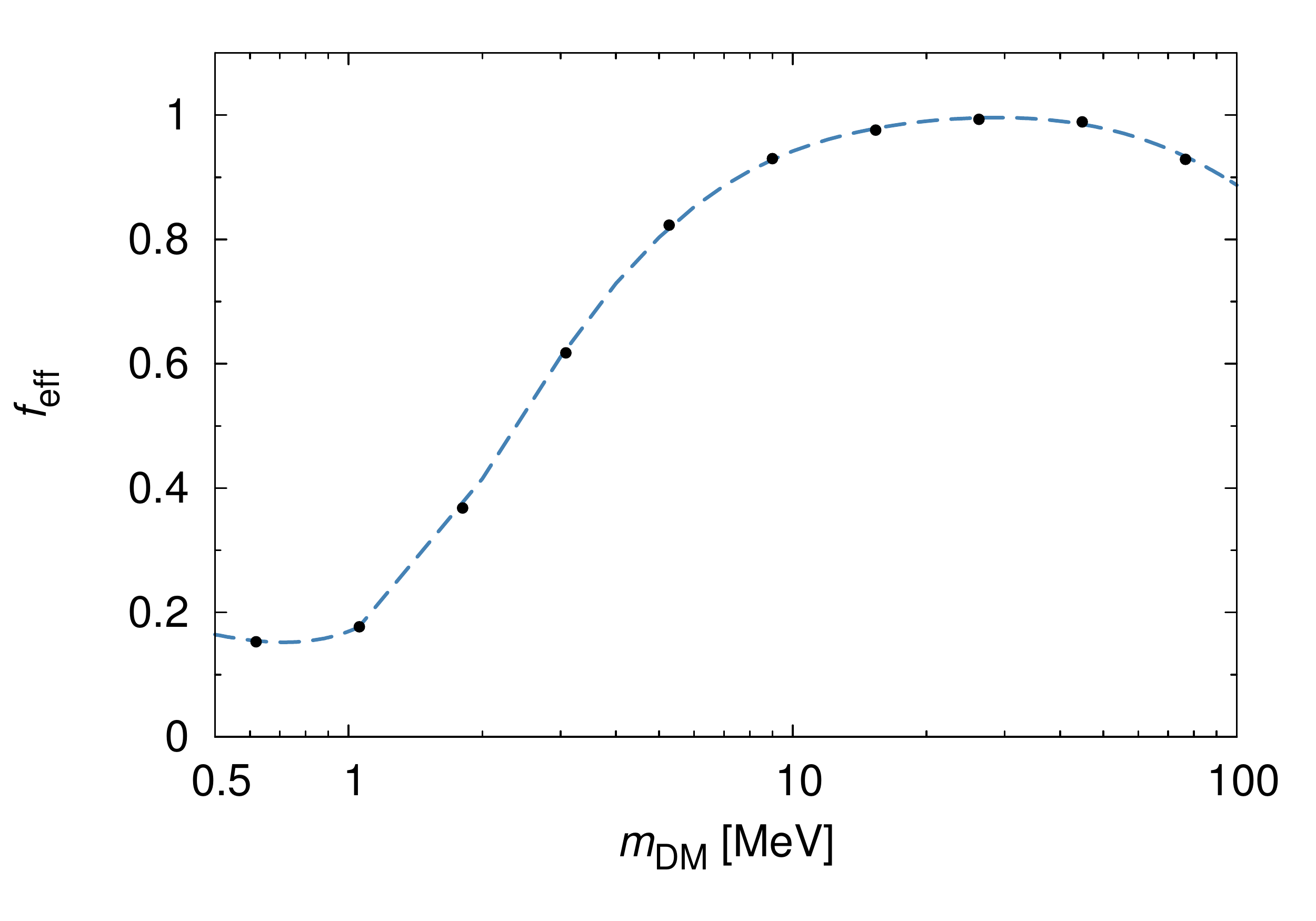}
\caption{The effective energy deposition fraction for the smooth DM background component $f_{\rm eff}$ versus the DM mass $m_{\rm DM}$ for the $e^+e^-$ annihilation channel. The points are taken from {\tt{\href{http://nebel.rc.fas.harvard.edu/epsilon}{nebel.rc.fas.harvard.edu/epsilon}}}~\cite{Slatyer:2015jla}.}
\label{fig:feff}
\end{figure}

Fig.~\ref{fig:feff} shows the energy deposition efficiency $f_{\rm eff}(\mx)$. At the low masses relevant to the 511 keV signal, energy absorption in the IGM actually becomes quite inefficient, leading to weaker constraints than for heavier WIMPs. This is because much of the energy lost by electrons to inverse Compton Scattering in this energy range ends up in photons that are below the 10.2 eV threshold to excite neutral hydrogen. These photons thus stream freely, leading to distortions of the CMB blackbody spectrum but no measurable effect on the ionisation of temperature of the IGM \cite{Slatyer:2015kla}.

\section{Electron Sector Thermal Production}
\label{sec:electrons}

Given the strong constraints in the neutrino sector, it makes sense to examine the alternative scenario of thermal production entirely through $e^+ e^-$ annihilation. To accomplish this, the annihilation cross section must be suppressed at late times. A p-wave term, which can be obtained by e.g. the exchange of a $Z'$ mediator \cite{Boehm:2002yz}, can lead to such a suppression, proportional to the velocity squared: $\svee = a + bv^2$.

Assuming $bv^2 \simeq 3 \times 10^{-26}$ cm$^3$ s$^{-1}$ at freeze-out, the velocity-suppressed p-wave term is too low by over an order of magnitude to reproduce the 511 keV signal. This means that the constant $a \sim 10^{-30}$ cm$^3$ s$^{-1}$ term is still required. The dark age constraints on the neutrino sector scenario therefore also apply directly to $a$. However, at present, CMB limits cannot say anything about $b$ due to the low thermal velocities after recombination~\cite{Diamanti:2013bia}.

Rather than increasing the energy density in the neutrino sector as it becomes non-relativistic, a coupling to electrons leads light DM to transfer entropy into the \textit{visible} sector \cite{Ho:2012ug}. Fixing $\rho_\gamma$ to the observed value, this translates to an effective decrease of entropy in the neutrino sector and thus a lower $\Neff$. In contrast with the previous case, this gives rise to an increase in $Y_{\rm P}$ but to a \textit{lower} D/H, owing to the different evolution of the baryon-to-photon ratio~$\eta$~\cite{Nollett:2013pwa}.

Analogously to Eq. \eqref{eq:neffcmb}, the value of $\Neff$ at recombination ($\mx \gg T$) becomes:
\begin{equation}
N_{\rm eff}^{{\rm Equil,} e} \simeq 3.046 {\left[1 + \frac{g_{\rm DM}}{2} \frac{7}{22} {F(y_\nu|_{T_{{\rm dec},\nu}})} \right]}^{-4/3}~,
\label{eq:neffcmb2}
\end{equation}
i.e., one obtains a reduction in the relative energy density of the neutrino sector, leaving with an overall lower radiation component of the universe. Once more, this is shown in Fig.~\ref{fig:neff}.

We neglect DM--electron scattering during recombination as the scattering cross section would need to be significantly larger than the annihilation cross section to have a noticeable effect on the CMB acoustic peaks~\cite{Chen:2002yh,Dvorkin:2013cea}.

\section{New constraints on light WIMPs}

\begin{figure*}[t]
\begin{tabular}{l l l l}
\hspace{-4ex}
\includegraphics[width=0.25\textwidth,clip,trim=5ex 0ex 29ex 0ex]{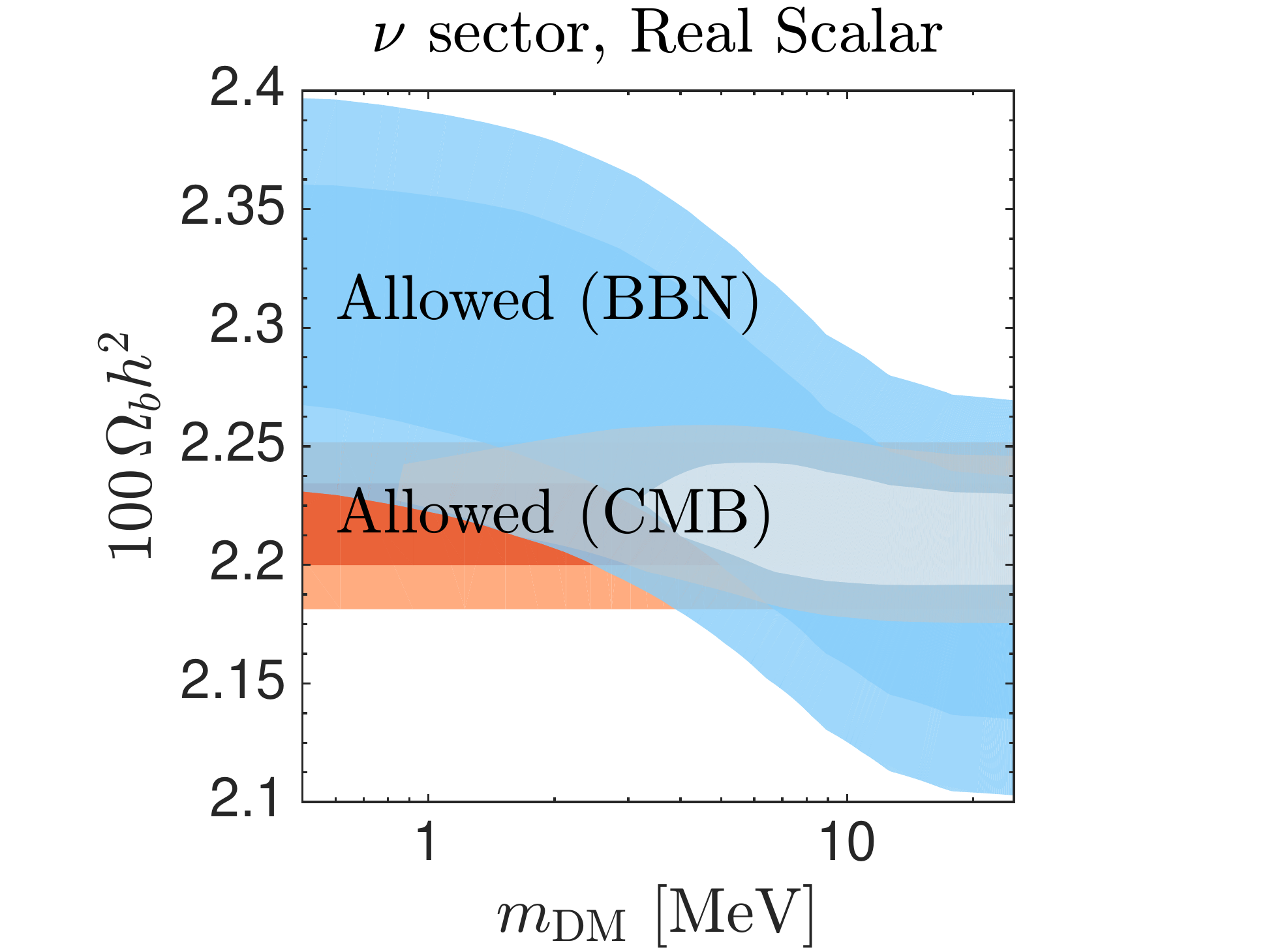} &
\hspace{-1.5ex}
\includegraphics[width=0.25\textwidth,clip,trim=5ex 0ex 29ex 0ex]{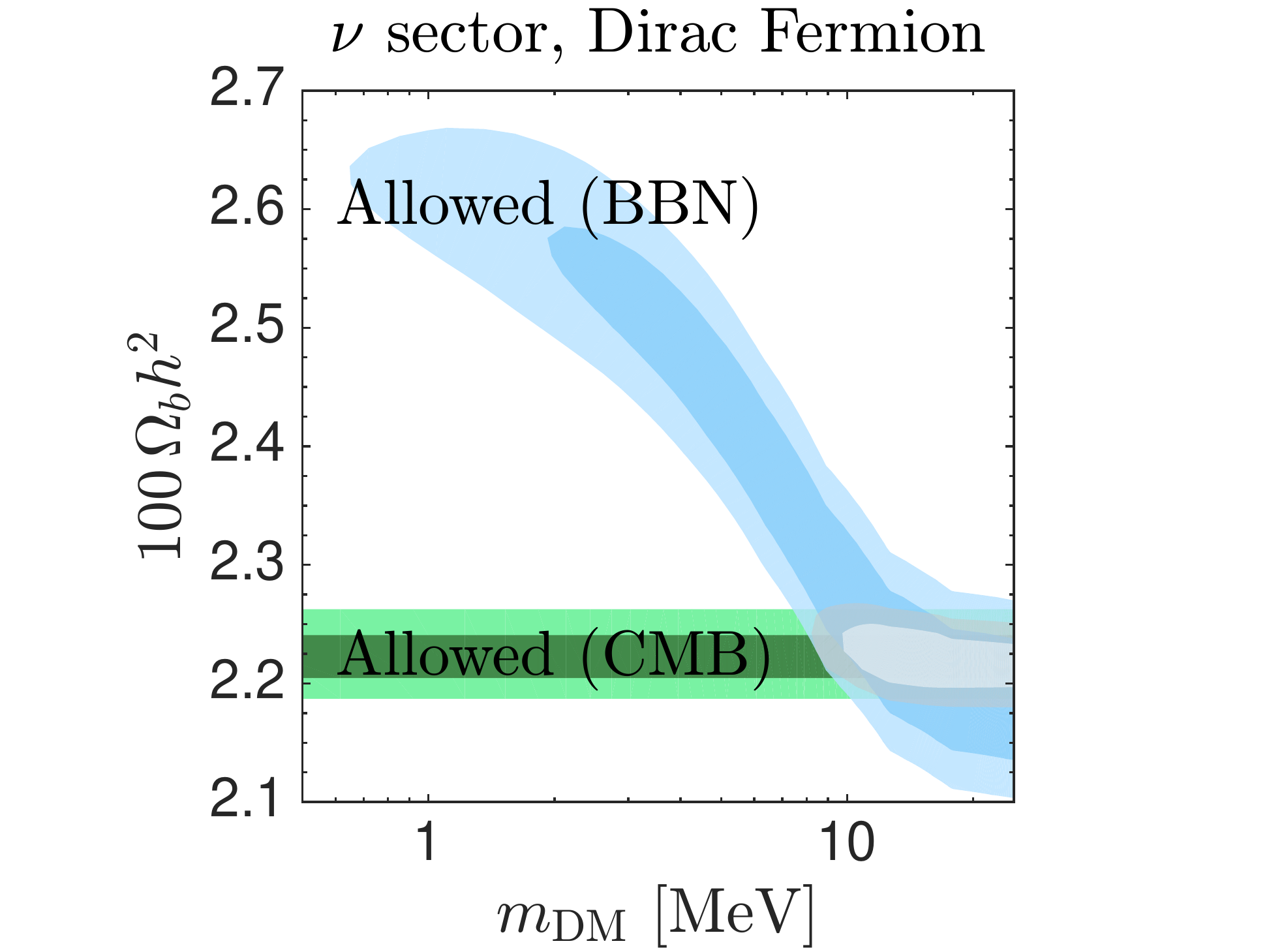}&
\hspace{-1.5ex}
\includegraphics[width=0.25\textwidth,clip,trim=5ex 0ex 29ex 0ex]{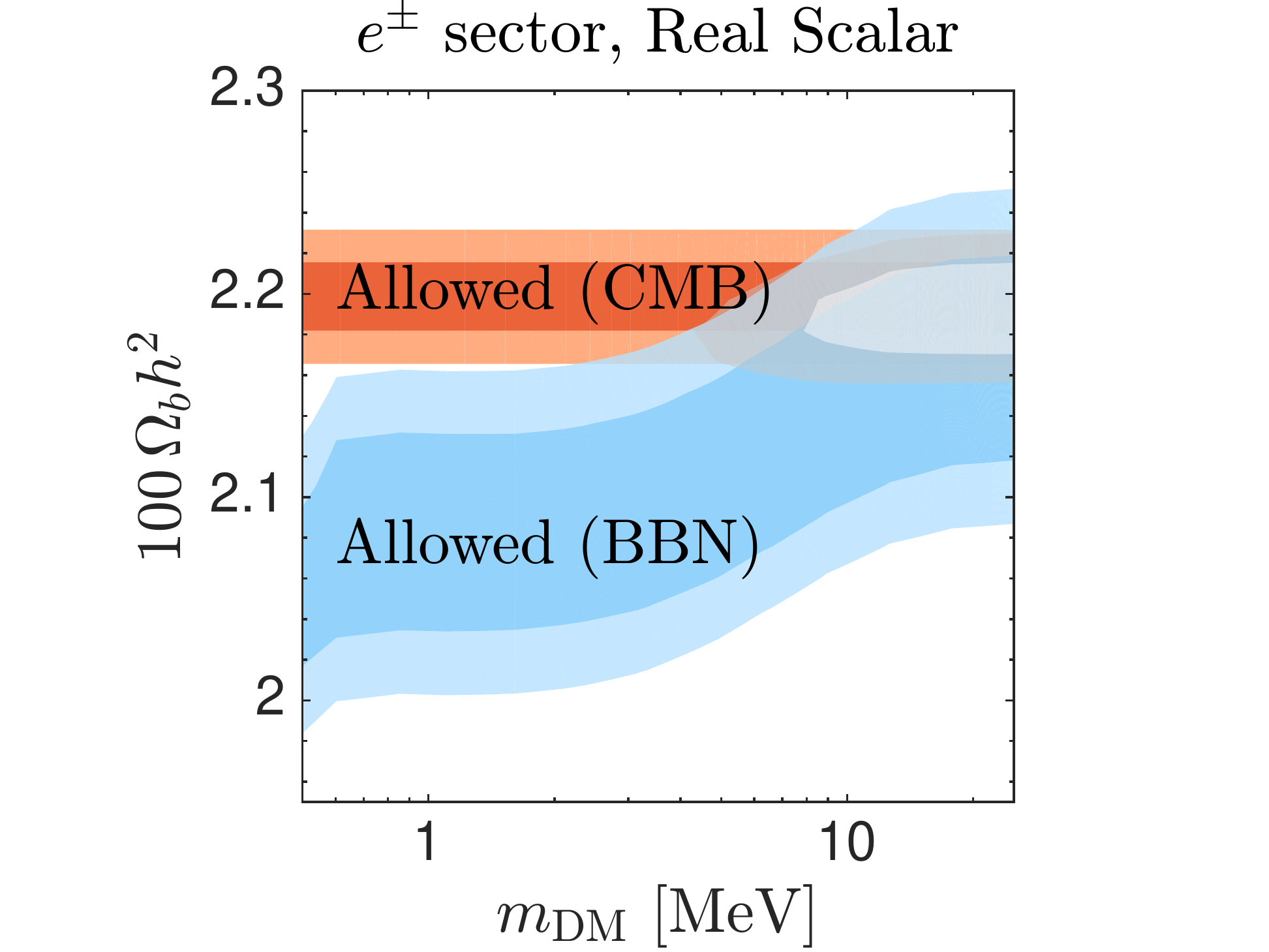}&
\hspace{-1.5ex}
\includegraphics[width=0.25\textwidth,clip,trim=5ex 0ex 29ex 0ex]{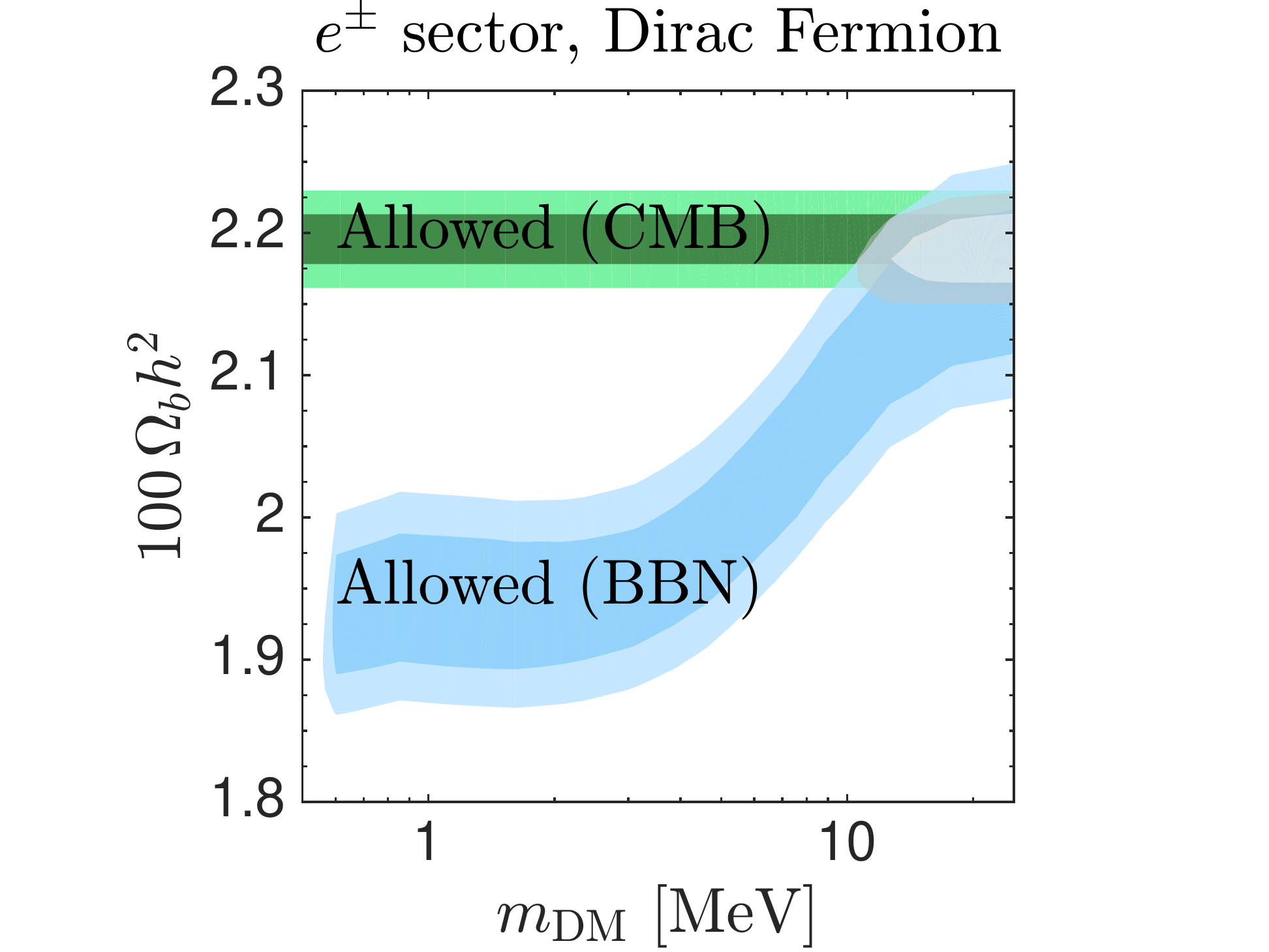} \\
\end{tabular}
\caption{Constraints on the baryon content $\Omega_{\rm b} h^2$ versus the light DM mass $m_{\rm DM}$ for the four considered scenarios. In orange/green, 68\% and 95\% CL regions allowed by \Planck; in blue, 68\% and 95\% CL allowed regions from direct measurements of $Y_{\rm P}$ and D/H. Only overlapping regions shown in grey are compatible with both datasets. BBN requirements on a Dirac fermion are in tension with the restriction that $\mx \lesssim 7$ MeV to avoid overproduction of bremsstrahlung gamma-rays~\cite{Ascasibar:2005rw, Beacom:2005qv,Sizun:2007ds}. An extensive MCMC analysis of CMB data is necessary to firmly rule out all possibilities (see Fig. \ref{fig:summary}).}
\label{fig:BBN}
\end{figure*}

In order to self-consistently evaluate the effects of each of these scenarios and predict the resulting CMB angular power spectra, the physics described in Secs.~\ref{sec:neutrinos} and \ref{sec:electrons} must be embedded into a CMB code that also accounts for a full recombination calculation. \Planck measurements of the temperature and polarisation angular power spectra already constrain extra ionisation, damping, and modifications of the universe's radiation content to unprecedented accuracy in the \lcdm model. We thus confront the results of the Boltzmann code {\sc class} with the data from \Planck, where we include DM--neutrino scattering (where applicable), in addition to the changes in $\Neff$ as a function of the DM mass, and the effect of energy injection in the dark ages due to ongoing DM self-annihilation.

To account for changes in the BBN era, we include in {\sc class} the modified $Y_{\rm P}$ due to light DM. To this end, we modify the {\sc PArthENoPE}~\cite{Pisanti:2007hk} code to compute $Y_{\rm P}$ and D/H for arbitrary $m_{\rm DM}$, $\Omega_b h^2$ pairs. We also update the $d(p,\gamma)^3$He, $d(d,n)^3$He and $d(d,p)^3$H reaction rates in {\sc PArthENoPE} with more precise determinations \cite{Coc:2015bhi}, and take a fixed neutron lifetime $\tau_n=880.3$~s~\cite{Agashe:2014kda}. 

For each scenario, we perform a Markov Chain Monte Carlo (MCMC) search using the {\sc Monte Python}~\cite{Audren:2012wb} code. This is in contrast with Refs.~\cite{Nollett:2013pwa,Steigman:2014pfa, Nollett:2014lwa}, who compared predicted changes in $N_{\rm eff}$ directly with derived \lcdm parameters from \Planck. By recomputing the full recombination history and comparing directly with the measured power spectra, we are able to fully account for the effect of degeneracies between cosmological parameters.

The MCMC searches include the six base \lcdm parameters ($H_0, \Omega_{\rm DM} h^2, \Omega_{\rm b} h^2, A_{\rm s}, n_{\rm s}, \tau_{\rm reio}$). In the neutrino sector scenario, we add the DM mass $\mx$, the energy injection rate $p_{\rm ann}$ and a parameterisation of the DM--neutrino scattering cross section $u \propto \sigma_{\mathrm{DM}-\nu}$. $u$ must be marginalised (integrated) over, along with the \lcdm parameters. In the electron sector case, the additional parameters are simply $\mx$ and $p_{\rm ann}$. We use the ``\Planck+ lensing'' 2015 dataset, which includes the latest TT, TE, EE and lowP data~\cite{Adam:2015rua}. The addition of BAO, supernovae data and an $H_0$ HST prior do not significantly change our posterior distributions.

\begin{figure*}[t]
\centering
\includegraphics[width=0.35\textwidth,angle=-90,clip,trim=0ex 0ex 0ex 5ex]{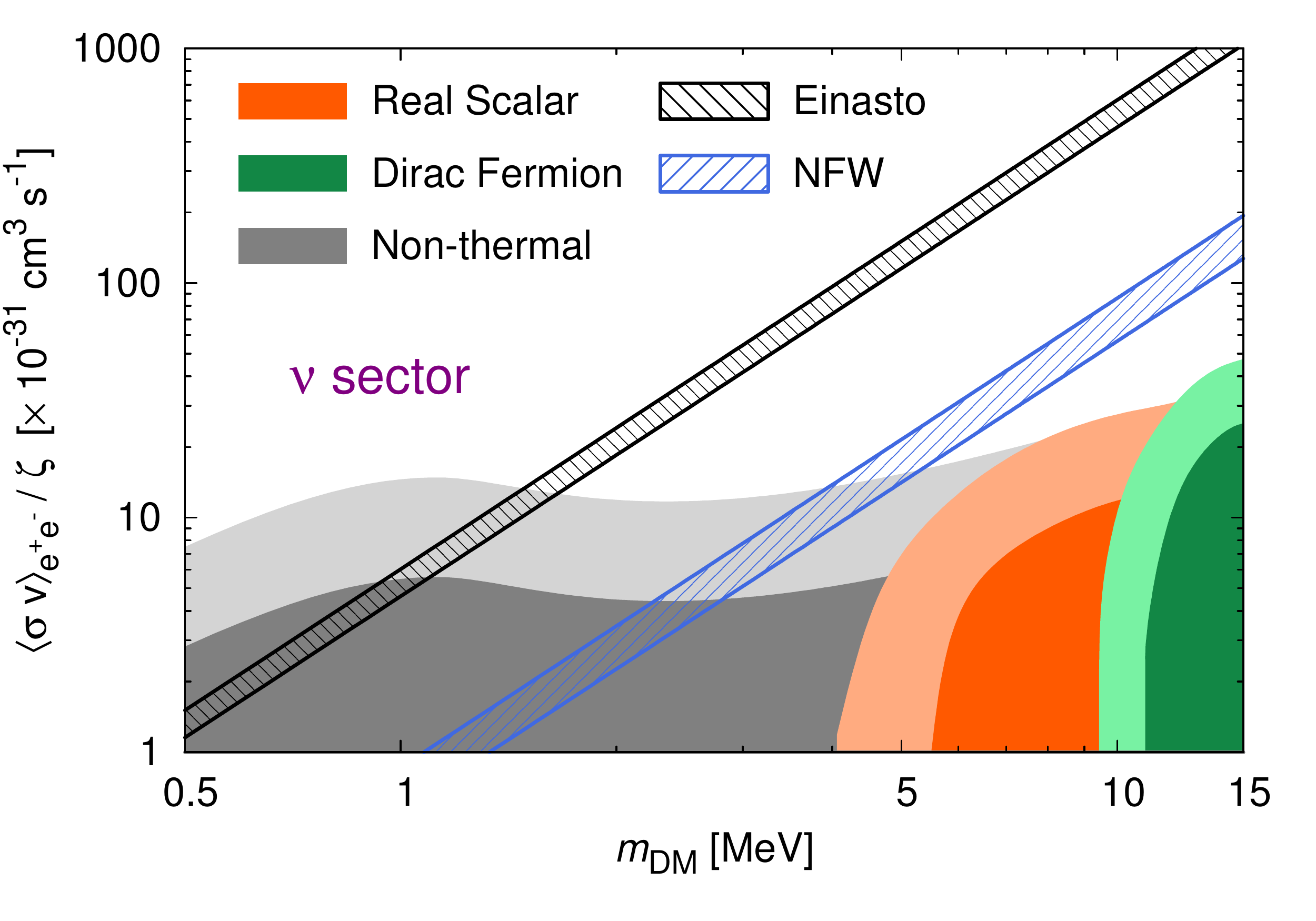}
\hspace{2ex}
\includegraphics[width=0.35\textwidth,angle=-90,clip,trim=0ex 0ex 0ex 5ex]{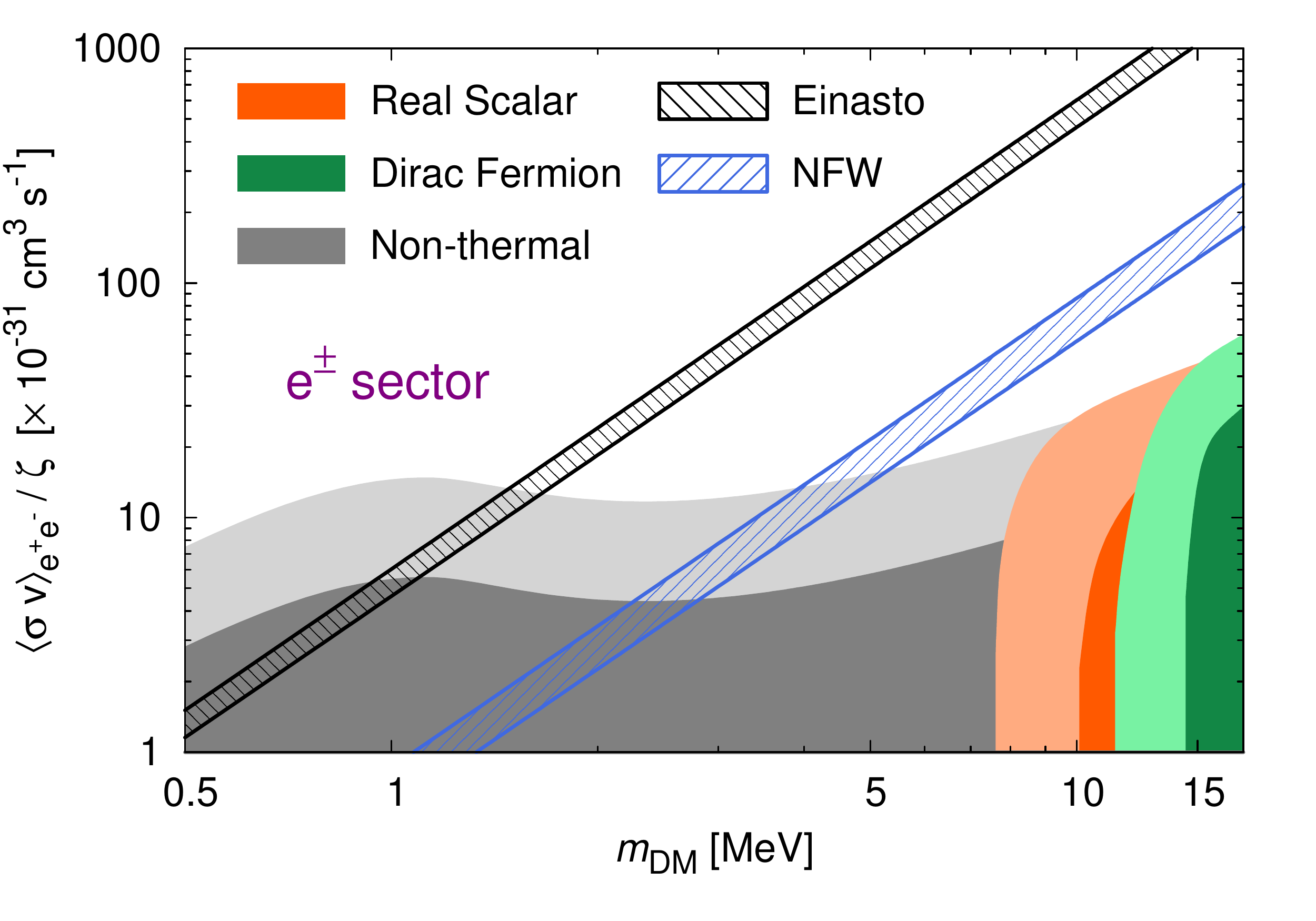}
\vspace{-2ex}
\caption{The DM annihilation cross section into $e^+ e^-$ as a function of the mass of the DM particle. $\zeta =1$ when the DM and its antiparticle are identical, and 1/2 otherwise. Hatched bands show the  values of $\svee$ vs. $m_{\rm DM}$ that are necessary to explain the 511 keV line for Einasto (black, upper) and NFW (blue, lower) DM density profiles, including the $\pm 2\sigma$ uncertainty from the DM flux, halo shape and stellar disk component~\cite{Vincent:2012an}. In both panels, values of $\svee$ above the grey allowed regions are excluded by \Planck CMB limits on energy injection in the dark ages~\cite{Adam:2015rua}. The coloured contours correspond to the 68\% and 95\% CL regions that are allowed by \Planck CMB data for thermal production via the neutrino sector (left panel) and electron sector (right panel); we consider a real scalar WIMP (orange) and a Dirac fermion WIMP (green). Bounds on the DM mass from the entropy transfer [Eqs.~\eqref{eq:neffcmb} and~\eqref{eq:neffcmb2}] constrain the coloured regions from the left, while bounds from late-time energy injection on $\svee$ constrain them from above. The combination of these effects allows us to rule out the DM mass range required to explain the 511 keV line.}
\label{fig:summary}
\end{figure*}

Before turning to our main results, we first follow the approach of Refs.~\cite{Boehm:2012gr, Boehm:2013jpa,Nollett:2013pwa,Steigman:2014pfa, Nollett:2014lwa} and show constraints from direct measurements of $Y_{\rm P}$ and D/H based on changes during BBN, employing the recommended PDG determinations~\cite{Agashe:2014kda}:
\begin{eqnarray}
\mathrm{D/H} &=& (2.53 \pm 0.04) \times 10^{-5}~; \nonumber \\
Y_{\rm P} &=& 0.2465 \pm 0.0097~. \nonumber
\end{eqnarray}
We include a 2\% theory error on our D/H calculation, while the experimental error on $Y_{\rm P}$ is dominant \cite{Coc:2015bhi}. We note that previous studies have used a higher determination of $Y_{\rm P} = 0.254 \pm 0.003$~\cite{Izotov:2013waa}. This value is incompatible with the best fit \lcdm parameters obtained by the \Planck experiment at more than 3$\sigma$. However, when it is combined with our CMB analysis, it has very little effect on our mass bounds. We thus use the recommended PDG value given above.

The 68\% and 95\% CL allowed regions are shown as blue bands in Fig.~\ref{fig:BBN}. Horizontal bands show the allowed 68\% and 95\% CL posterior regions for $\Omega_{\rm b} h^2$ from \Planck data for a real scalar WIMP (orange) and a Dirac fermion WIMP (green). The other possibilities (complex scalar, Majorana fermion or vector) would be more constrained than the real scalar case. For clarity, we do not show them.

In each case, only the overlapping regions shown in grey are allowed. Therefore, $m_{\rm DM} \gtrsim 8$ MeV is required for Dirac DM, in conflict with the spectral constraints ($m_{\rm DM} \lesssim 7$ MeV) from INTEGRAL/SPI observations~\cite{Ascasibar:2005rw, Beacom:2005qv, Sizun:2007ds}. In the real scalar case, this restriction is relaxed to $m_{\rm DM} \gtrsim 4$ MeV (electron sector) and $m_{\rm DM} \gtrsim 0.8$ MeV (neutrino sector).

The contours in Fig.~\ref{fig:BBN} are in general agreement with those presented in Refs.~\cite{Nollett:2013pwa,Steigman:2014pfa, Nollett:2014lwa} for a Majorana fermion DM particle, bearing in mind the updated BBN and CMB data used in our analysis. While  Fig.~\ref{fig:BBN} gives an indication of the combined power of CMB and BBN constraints, our MCMC scan using CMB observables alone provides the most robust exclusions, especially given the significant differences between primordial abundance measurements. We therefore turn to these results.

Fig.~\ref{fig:summary} shows the marginalised posterior limits from our MCMC for each scenario, compared with the cross section required to explain the 511 keV line with an annihilating WIMP. The hatched bands show the values of $\svee$ ($= a$ in the electron sector case) that fit the 511 keV intensity and morphology, including the $\pm 2\sigma$ uncertainty from the DM flux, halo shape and stellar disk component~\cite{Vincent:2012an}. The upper black band shows the best-fit region for an Einasto DM profile; the corresponding band for an NFW profile, which gives a significantly worse fit to the signal's morphology, is shown below it, in blue.

The grey contours show the 68\% and 95\% CL constraints on $\svee$ alone, due to ionisation of the IGM as described in Eq. \eqref{eq:dfdt}. The shape of these contours is due to the mass-dependence of $f_{\rm eff}$ (see Fig.~\ref{fig:feff}), leading to the requirement that $m_{\rm DM} \lesssim 1.5$ MeV (Einasto) and $m_{\rm DM} \lesssim 5$ MeV (NFW) at 95\% CL to explain the signal. This constraint is compatible with the most recent limit on $p_{\rm ann}$ given by the \Planck collaboration \cite{Planck:2015xua}. These bounds are independent of the relic density requirement, which we apply next, and therefore, directly constrain both thermal and non-thermal DM.

In both the neutrino and electron scenarios, the regions allowed by \Planck CMB observations (shown in orange and green) lie at DM masses and cross sections into $e^\pm$ that are respectively too heavy and too weak to reproduce the INTEGRAL/SPI signal. In all cases, the required annihilation rate to produce the positronium signal is outside the 99\%~CL (3$\sigma$) containment region. 
 
In the neutrino sector case, the lower bound at 95\% CL on the WIMP mass between 4 and 10 MeV (for $g_{\rm DM} \in \{1,4\}$) is mainly due to the high sensitivity of \Planck at larger multipoles to changes in $N_{\rm{eff}}$ and $Y_{\rm P}$\footnote{Note that these constraints would be slightly stronger if we had not marginalised over the DM--neutrino scattering parameter $u$.}. In the electron sector, these effects yield an even stronger bound, between 7 and 11 MeV at 95\% CL. Combined with the constraints on $p_{\rm ann}$ that limit the allowed regions from above, our results show that a light self-annihilating WIMP cannot be responsible for the 511 keV galactic line without severe disagreement with CMB data.

\section{Conclusions}

The WIMP hypothesis requires an origin of the relic density of dark matter (DM) via thermal freeze-out in the early universe. To simultaneously reproduce the galactic 511 keV line from positronium annihilation, the remaining branching fraction must be ``hidden'' from galactic and cosmological constraints. We have shown that the two methods of accomplishing this are insufficient: i) thermal production via the neutrino sector which, although invisible today, leads to a radiation component that is too large for early universe observables; or ii) p-wave (velocity-suppressed) production via the electromagnetic sector, giving too large of a \textit{reduction} in the universe's radiation content. 

Other scenarios exist; for example, eXciting dark matter (XDM) has been explored in depth~\cite{Finkbeiner:2007kk, Pospelov:2007xh, ArkaniHamed:2008qn, Finkbeiner:2009mi,Chen:2009av, Cline:2010kv, Bai:2012yq, Cline:2012yx} as an alternative mechanism to evade the suppressed self-annihilation cross sections. As pointed out by Ref. \cite{Frey:2013wh}, our dark ages constraints can also be applied to XDM; their forecasts show that \Planck should rule out XDM models with a mass splitting larger than $\sim 1.5$ MeV. Smaller splittings are possible but require tuning of the DM model.

We also note that one can mitigate the effects of entropy transfer and late-time energy injection by adding an extra component of dark radiation, or an extra source of photons or neutrinos between the epoch of neutrino decoupling and recombination. Such a coincidence would weaken our constraints; however, this type of model-building goes beyond the scope of our analysis.

The favoured DM explanation of the galactic 511 keV line, an anomaly that has endured for over four decades, is thus in fundamental disagreement with the latest precision cosmological data in the most ``vanilla'' of models, i.e. thermal production with no extra particles. As the origin of the positrons in the galactic bulge remains unknown, an alternative DM model may yet be responsible; however, the light WIMP hypothesis is no longer viable.

\begin{acknowledgments}

The authors thank O.~Mena and M.~Escudero for useful discussions. RJW is supported by the STFC grant ST/K501979/1. This work was supported by the European Union FP7 ITN INVISIBLES (Marie Curie Actions, PITN-GA-2011-289442). The work of CM is part of the research programme of the Foundation for Fundamental Research on Matter (FOM), which is part of the Netherlands Organisation for Scientific Research (NWO). CM thanks SURFsara for the use of the Lisa Computer Cluster. ACV is
supported by an Imperial College London Junior Research Fellowship.

\end{acknowledgments}

%

\end{document}